# DISORDER EFFECTS IN *d*-WAVE SUPERCONDUCTORS

C. T. Rieck, K. Scharnberg (`scharnbe@physnet.uni-hamburg.de`),
and S. Scheffler
*I. Institut für Theoretische Physik, Universität Hamburg*
*20355 Hamburg, Germany*

**Abstract.**
In the theoretical analyses of impurity effects in superconductors the assumption is usually made that all quantities, except for the Green functions, are slowly varying functions of energy. When this so-called Fermi Surface Restricted Approximation is combined with the assumption that impurities can be represented by $\delta$-function potentials of arbitrary strength, many reasonable looking results can be obtained. The agreement with experiments is not entirely satisfactory and one reason for this might be the assumption that the impurity potential has zero range. The generalization to finite range potentials appears to be straightforward, independent of the strength of the potential. However, the selfenergy resulting from scattering off finite range impurities of infinite strength such as hard spheres, diverges in this approximation at frequencies much larger than the gap amplitude! To track down the source of this unacceptable result we consider the normal state. The elementary results for scattering off a hard sphere, including the result that even an infinitely strong $\delta$-function potential does not lead to scattering at all in systems of two and more dimensions, are recovered only when the energy dependencies of all quantities involved are properly taken into account. To obtain resonant scattering, believed to be important for the creation of mid-gap states, the range of the potential is almost as important as its strength.

## 1. Introduction

Scattering of a particle by a potential is a time-honored problem of quantum mechanics. (Taylor, 1972) It is straightforward to formulate although a microscopic derivation of the scattering potential requires very intricate considerations. Most often the scattering potential is modelled by some plausible function in real space. This is the route we shall follow here.

Even though potential scattering is elastic, calculation of the scattered wave function requires consideration of all virtual states, up to infinitely high energies. When the scattered wave function is calculated for piecewise constant potentials by imposing boundary conditions, the need to include high energy virtual states is not apparent. We shall not solve the Schrödinger equation directly but rather use the Green function formalism, because this allows a straightforward generalization to scattering in a metal containing an ensemble of defects, even when the metal goes superconducting. However, in order to check the approximations usually made in the application of this formalism we shall revisit the simplest problems of scattering theory and reproduce known results.

The driving force behind the study of disorder effects in superconductors is the hope that such investigations will give information on the pairing state and the pairing interaction. Indeed, qualitative differences are expected between conventional and unconventional superconductors, the latter being defined by a vanishing Fermi surface average of the order parameter. Potential scattering in an anisotropic conventional superconductor would at high enough concentration lead to a finite, isotropic gap, while unconventional superconductors are expected to acquire midgap states before superconductivity is destroyed. Conventional superconductors show this kind of behavior in the presence of spin-flip scattering. (Abrikosov and Gor'kov, 1961) Because of the innate magnetism in high temperature superconductors, nonmagnetic impurities can induce local moments and thus blur the seemingly clear distinction between potential and spin-flip scattering. Here, we shall assume that we are dealing with $d$-wave superconductors. Since for unconventional superconductors there is no qualitative difference between these two types of scattering, we shall confine ourselves to the study of potential scattering. Even with this limitation there is a wide range of theoretical predictions as regards $T_c$-suppression, density of states, transport properties etc, depending on the electronic structure, that is assumed, the way disorder is modelled and depending on the analytical and numerical approximations employed. (Atkinson et al., 2000; Hirschfeld and Atkinson, 2002; Balatsky et al., 2006)

In a solid the scattering potential is due to defects of the crystalline lattice which are distributed more or less randomly over the whole sample. For large systems, an average over defect configurations has to be taken, not only because the problem would be untractable otherwise but also because the defect configuration is unknown and, except at very low temperatures, changes with time. Thus the problem of treating disorder in solids is usually broken down into two parts: scattering off a single defect and averaging with respect to such individual scattering events. As model for disorder we shall use an alloy model in which some of the host atoms are randomly replaced by some other kind of atom. Averaging independently with respect to all possible defect positions, which limits the applicability of this theory to small concentration of defects, leads to the selfconsistent $T$-matrix approximation (SCTMA).

We shall use the $T$-matrix equation (Lippmann-Schwinger equation (Taylor, 1972)) to describe an individual scattering event. This is a two-dimensional Fredholm integral equation of the second kind with a singular kernel. When the scattering potential is so weak that this equation can be solved by iteration up to 2nd order (Born approximation), integrals are not actually evaluated but are parame-trized through quantities like lifetimes, transport times, and shifts in the chemical potential. Then, a detailed knowledge of the potential is necessary only when a microscopic calculation of these parameters is undertaken. We are interested in the case of strong scatterers, including infinitely high potentials, for which the Born series diverges. The limiting case might seem unphysical but we know from elementary quantum mechanics that scattering off a hard sphere is described by perfectly well-behaved wave functions. The reason for the wide spread interest in the strong scattering limit is the realisation that in systems with energy gaps strong repulsive potentials might create states inside the gaps. (Joynt, 1997; Balatsky et al., 2006)

From this $T$-matrix the local density of states (LDOS) can be obtained. The selfenergy of the configuration averaged and hence translationally invariant Green function is given by a different but closely related quantity which might be called generalized $T$-matrix. (Taylor, 1972; Mahan, 1981) The Green function appearing in the kernel of the integral equation which determines this generalized $T$-matrix depends on some arbitrary energy $\omega$. Only when this is set equal to the energy of the scattered particle does one recover the $T$-matrix known from scattering theory.

The non-iterative solution of the two-dimensional integral equation, which gives the desired $T$-matrix, is a formidable problem which we have not yet solved!

One frequently used simplification, especially in the theory of superconductivity, is to fix all momentum variables at the Fermi energy except in the Green function which is the only quantity integrated with respect to energy. (Fermi surface restricted approach) However, this energy integral diverges! This is related to the fact that the real part of the Green function in position space $G_\omega(\mathbf{r}, \mathbf{r}')$ diverges for $\mathbf{r}' \to \mathbf{r}$ in two and three dimensions. The divergence can be removed by invoking particle-hole symmetry which eliminates the offensive term, (Flatté and Byers, 1999; Salkola et al., 1996) by averaging the Green function with respect to $\mathbf{r}'$ over a small volume around $\mathbf{r}$, (Flatté and Byers, 1999) or by restricting consideration to a single band of finite width. Omitting the divergent term is central to the immensely successful quasi-classical theory of superconductivity. (Eilenberger, 1968; Larkin and Ovchinnikov, 1969; Serene and Rainer, 1983) The quite convincing argument for this approach is that the divergence comes from energies far removed from the Fermi energy were the modifications of the electronic states due to the onset of superconductivity are negligible. So, when differences between superconducting and normal state properties are calculated one expects such contributions to cancel. In many cases, however, no such differences are calculated: one does expect the results for the superconducting state to reduce to their normal state equivalents as $T \to T_c$.

In much of the published work, the theory is further simplified by assuming $\delta$-function potentials. With this assumption, which has been made by many authors including ourselves (Hensen et al., 1997), the (generalized) $T$-matrix equation is no longer an integral equation and even for a superconductor the solution is trivial. Then a variety of interesting results can be derived, most of which are in reasonable agreement with experiment. Modelling the scattering potential by a $\delta$-function potential actually reduces the complexity of the problem to such an extent that it is possible to drop the assumption of scattering events at different defects being independent. (Atkinson et al., 2003) We should note, however, that according to elementary quantum mechanics a $\delta$-function potential in two and three dimensions does not scatter at all.

Even for defects within the $CuO_2$-planes it seems rather doubtful that their effect is limited to a single site. Defects due to oxygen nonstoichiometry and cation disorder, which reside on lattice sites away from the conducting $CuO_2$-planes, are only poorly screened and hence are certainly long ranged. One conceivable consequence of the finite range of the defect potentials is a mitigation of the $T_c$ suppression by potential scattering in unconventional superconductors. Balian *et al.*(Balian and Werthamer, 1963) already noted that forward scattering would reduce the deleterious effect of potential scattering on $T_c$, while Foulkes and Gyorffy presented a detailed calculation for *p*-wave superconductors in the Born approximation, showing that it is the transport time that controls the $T_c$-suppression. (Foulkes and Gyorffy, 1977) Millis *et al.* seem to have been the first to apply these ideas to *d*-wave superconductors, again using the Born approximation.(Millis et al., 1988)

We tried to generalize the treatment of angle dependent scattering to arbitrarily large potentials, still using the Fermi surface restricted approach so that only one-dimensional integral equations had to be solved. (Rieck et al., 2005) As model potential we used a Gaussian for computational convenience. We shall show below that in the unitary limit the normal state selfenergy calculated within the SCTMA is proportional to the number of scattering channels considered. For the pair breaking parameter a similarly unphysical result is found when the unitary limit is taken.

In Section 2 we develop the general theory of potential scattering in a *d*-wave superconductor within a continuum description of the electronic structure. In the following section we introduce the widely used Fermi surface restricted approximation and calculate the selfenergy and the pair breaking parameter using the selfconsistent *T*-matrix approximation. We show that these results in the strong scattering limit must be wrong and deduce from that the inadequacy of the Fermi surface restricted approximation, at least in the context of potential scattering in a metal. In Section 4 we study scattering off a single impurity and demonstrate what needs to be done in order to get reliable results.

## 2. Basic theory

The Gor'kov equations for the retarded Green functions describing a weak coupling spin singlett superconductor in the absence of magnetic interactions can be written in Nambu space as

$$\left(\omega_+\hat{\sigma}_0 - h(r, \nabla^2)\hat{\sigma}_3\right)\hat{G}(r, r'; \omega_+) - \int d^2\rho \Delta(r, \rho)\hat{\sigma}_1 \hat{G}(\rho, r'; \omega_+)$$
$$= \delta(r - r')\hat{\sigma}_0 \quad (1)$$

The $\hat{\sigma}$'s are Pauli matrices and $\hat{G}$ is a $2 \times 2$ matrix with only two independent elements. Though not strictly necessary, the introduction of a matrix Green function proves to be very useful. The Hamiltonian in (1) is given by

$$h(r, \nabla^2) = -\frac{1}{2\mu}\nabla^2 - \epsilon_F + V(r) + V_{\text{lattice}}(r). \quad (2)$$

$V(r)$ is a defect potential to be specified later and $V_{\text{lattice}}(r)$ is the periodic lattice potential. A complete solution for these Gor'kov equations is not in sight. One either uses a nearly free electron model, i.e. essentially ignores $V_{\text{lattice}}(r)$, or one uses a localized description requiring numerical calculations on a lattice of finite size. The relation between these two limiting cases will be discussed elsewhere. In an intermediate model some tight-binding dispersion relation leading to a band of finite width is introduced, but the difference between plane waves and Bloch functions is ignored. In this paper we treat the charge carriers as a two-dimensional (nearly) free electron gas. The Fermi surface in this model is circular and the band width is infinite and without $V(r)$ the system is translationally invariant. In particular, the order parameter in (1) will then depend only on $r - \rho$, so that (1) can easily be solved by Fourier transformation:

$$\hat{G}^0(k; \omega_+) = \frac{\omega\hat{\sigma}_0 + \varepsilon(k)\hat{\sigma}_3 + \Delta(k)\hat{\sigma}_1}{\omega_+^2 - \varepsilon^2(k) - \Delta^2(k)} = \sum_i G^{0i}(k, \omega_+)\hat{\sigma}_i. \quad (3)$$

True to our assumptions we should have $\varepsilon(k) = \frac{1}{2\mu}k^2 - \epsilon_F$ with $\mu$ some effective carrier mass, but one could also insert any (model) dispersion relation. The order parameter $\Delta$ is assumed to have *d*-wave symmetry with respect to its dependence on $k$. Using the Green function $\hat{G}^0$ one can rewrite the integro-differential equation (1) as inhomogeneous integral equation. However, the perturbation $V(r)$ affects the off-diagonal elements of $\hat{G}$ which, via the weak-coupling selfconsistency equation, change the order parameter, even when we ignore the possibility that the pairing interaction is modified by the presence of the defect(s). We, therefore, write the order parameter as

$$\Delta(r, \rho) = \Delta(r - \rho) + \delta\Delta(r, \rho). \quad (4)$$

The equation for $\hat{G}$ then has the general form of a Fredholm integral equation of the second kind

$$\hat{G}(\mathbf{r},\mathbf{r}') = \hat{G}^0(\mathbf{r},\mathbf{r}') + \int d^2\rho\, \hat{K}(\mathbf{r},\boldsymbol{\rho})\, \hat{G}(\boldsymbol{\rho},\mathbf{r}'). \tag{5}$$

With the order parameter fluctuations taken into account, the kernel

$$\hat{K}(\mathbf{r},\boldsymbol{\rho}) = \int d^2\rho'\hat{G}^0(\mathbf{r},\boldsymbol{\rho}')\{V(\boldsymbol{\rho})\delta(\boldsymbol{\rho}-\boldsymbol{\rho}')\hat{\sigma}_3 + \delta\Delta(\boldsymbol{\rho}',\boldsymbol{\rho})\hat{\sigma}_1\} \tag{6}$$

is itself an integral. Since the frequency only appears as parameter we have suppressed it for clarity. For an isotropic $s$-wave superconductor one has $\delta\Delta(\boldsymbol{\rho}',\boldsymbol{\rho}) = \delta\Delta(\boldsymbol{\rho})\,\delta(\boldsymbol{\rho}-\boldsymbol{\rho}')$. Then the suppression of the order parameter near a defect does not complicate the problem significantly.

For such a (3D) superconductor containing a single spherical impurity, Fetter (Fetter, 1965) has calculated the Friedel oscillations of the electron density and the order parameter amplitude in the asymptotic regime. The assumption of a zero-range pairing interaction is justified for conventional superconductors because the range of the pairing interaction is certainly much less than the coherence length considered to be the shortest length relevant for superconductivity. For cuprate superconductors even this assumption can be called into question because the coherence length is very short and the pairing interaction must have a finite range to allow for $d$-wave pairing. Starting from a weak-coupling version of the spin-fluctuation exchange (Dahm et al., 1993) one finds that the pairing interaction extends over several lattice constants and hence is comparable with the in-plane coherence length. Friedel oscillations occur on a length scale given by the Fermi wavelength. So, both for conventional and for cuprate superconductors the assumption, that the range of the pairing interaction is short compared with the length scale of the defect induced order parameter fluctuations, seems to be hard to justify.

Since in our view there are even more important shortcomings in the description of defects in unconventional superconductors, we shall neglect these order parameter fluctuations for the time being. Then Eqs. (5) and (6) reduce to

$$\hat{G}(\mathbf{r},\mathbf{r}';\omega) = \hat{G}^0(\mathbf{r}-\mathbf{r}';\omega) + \int d^2\rho\, \hat{G}^0(\mathbf{r}-\boldsymbol{\rho};\omega)\, V(\boldsymbol{\rho})\hat{\sigma}_3\, \hat{G}(\boldsymbol{\rho},\mathbf{r}';\omega) \tag{7}$$

Since we are interested in very strong potentials for which the Born series does not converge, the solution of this equation for general $V(\boldsymbol{\rho})$ is still quite a tall order, especially as this (screened) potential ought to be calculated selfconsistently taking into account the Friedel oscillations it induces in the charge density. Again, we neglect this effect and use some model potential.

The most popular model is a $\delta$-function potential

$$V(\boldsymbol{\rho}) = V\,\delta(\boldsymbol{\rho}), \tag{8}$$

because for this potential the task of solving (7) is trivial:

$$\hat{G}(r, r'; \omega) = \hat{G}^0(r - r'; \omega) + V \hat{G}^0(r; \omega) \hat{\sigma}_3 \left( \hat{\sigma}_0 - V \hat{G}^0(0; \omega) \hat{\sigma}_3 \right)^{-1} \hat{G}^0(-r'; \omega) \quad (9)$$

Unless regularized as described in the introduction, the diagonal components of $\hat{G}^0(0; \omega) = \int \frac{d^D k}{(2\pi)^D} \hat{G}^0(k; \omega_n)$ diverge in two and three dimensions. Hence

$$\hat{G}(r, r'; \omega) = \hat{G}^0(r - r'; \omega) \quad (10)$$

and we have to conclude that $\delta$-function potentials, no matter how strong, have no effect on the properties of a superconductor. This is in accord with the results, obtainable by elementary quantum mechanics, for the scattering phase shifts of sphere- or disk-shape potentials (see Figs. 3 and 4). When the divergence is removed by hand, $\hat{G}^0(r - r'; \omega)$ no longer solves the original equation of motion (1).

The order of the divergence can be seen from the explicit forms of the Green's functions in position space, which are easily calculated using an integral representation for the Hankel function (Watson, 1952), when the order parameter in (3) is momentum independent. Since the result for the two-dimensional case is not readily available (Scheffler, 2004), we shall give it here for $\omega > 0$:

$$G^{00}(r, \omega_+) = -\frac{m}{4} \frac{\omega_+}{\sqrt{\omega_+^2 - \Delta^2}} [iJ_0(r\Omega_+) + iJ_0(r\Omega_-) - Y_0(r\Omega_+) + Y_0(r\Omega_-)] \quad (11)$$

$$G^{03}(r, \omega_+) = -\frac{m}{4} [iJ_0(r\Omega_+) - iJ_0(r\Omega_-) - Y_0(r\Omega_+) - Y_0(r\Omega_-)] \quad (12)$$

with $\quad \Omega_\pm = k_F^2 \pm 2m\sqrt{\omega_+^2 - \Delta^2} \quad (13)$

$J_0$ and $Y_0$ are Bessel functions of the first and second type. For small argument, the leading term in $Y_0(z)$ is $0.5\pi \ln(0.5z)$. Hence, $\mathcal{R}e G^{03}(r, \omega_+)$ diverges as $\ln r$ for $r \to 0$, while the divergent terms in $G^{00}$ cancel.

If corresponding expressions were available for *d*-wave superconductors it would probably be easiest to calculate the local density of states in the vicinity of a single defect for given $V(\rho)$ directly from (7). Since this is not the case and in view of the interest in properties of superconductors containing random ensembles of defects, we use an eigenfunction represention for $\hat{G}^0(r - r'; \omega)$ and introduce a generalized *T*-matrix to rewrite (7) as:

$$\hat{G}(r, r', \omega) = \hat{G}^0(r - r', \omega) + \int \frac{d^2 k}{(2\pi)^2} \int \frac{d^2 k'}{(2\pi)^2} e^{ikr} \hat{G}^0(k, \omega) \hat{T}(k, k'; \omega) \hat{G}^0(k', \omega) e^{-ik'r'} \quad (14)$$

$$\hat{T}(k, k'; \omega) = V(k - k')\hat{\sigma}_3 + \int \frac{d^2 p}{(2\pi)^2} V(k - p) \hat{\sigma}_3 \hat{G}^0(p, \omega) \hat{T}(p, k'; \omega) \quad (15)$$

One possible description of an ensemble of (identical) impurities is the so-called alloy model

$$V(\mathbf{r}) = \sum_{i=1}^{N} v(\mathbf{r} - \mathbf{R}_i) \qquad N \gg 1. \tag{16}$$

Taking an average with respect to the random impurity sites $\mathbf{R}_i$, neglecting interference between scattering processes at different sites, consistent with assuming a small concentration $n_{\text{imp}}$ of impurities, gives

$$\hat{t}(\mathbf{k}, \mathbf{k}'; \omega) = v(\mathbf{k} - \mathbf{k}')\hat{\sigma}_3 + \int \frac{d^2 p}{(2\pi)^2} v(\mathbf{k} - \mathbf{p}) \, \hat{\sigma}_3 \, \hat{G}(\mathbf{p}, \omega) \, \hat{t}(\mathbf{p}, \mathbf{k}'; \omega) \tag{17}$$

This equation is very similar to (15): $\hat{t}$ is now the $\hat{T}$-matrix for a single defect and $\hat{G}^0(\mathbf{k}, \omega)$ has been replaced by

$$\hat{G}(\mathbf{k}, \omega) = \left[ \omega\hat{\sigma}_0 - \varepsilon(\mathbf{k})\hat{\sigma}_3 - \Delta(\mathbf{k})\hat{\sigma}_1 - \hat{\Sigma}(\mathbf{k}, \omega) \right]^{-1} = \sum_i G^i(\mathbf{k}, \omega) \, \hat{\sigma}_i. \tag{18}$$

which, like $\hat{G}^0(\mathbf{k}, \omega)$, describes a translationally invariant system.

$$\hat{\Sigma}(\mathbf{k}, \omega) = n_{\text{imp}} \, \hat{t}(\mathbf{k}, \mathbf{k}; \omega) \tag{19}$$

is a selfenergy which has to be calculated selfconsistently. When $\hat{t}(\mathbf{k}, \mathbf{k}'; \omega)$, together with the selfenergy, are expanded in terms of Pauli matrices:

$$\hat{t} = t^0 \hat{\sigma}_0 + t^1 \hat{\sigma}_1 + i t^2 \hat{\sigma}_2 + t^3 \hat{\sigma}_3 \tag{20}$$

one obtains four coupled 2D integral equations for $t^\ell$, $\ell = 0, \ldots, 4$

$$t^0 = \int \frac{d^2 p}{(2\pi)^2} v \left[ \frac{\omega - \Sigma^0}{D} t^3 + \frac{\varepsilon + \Sigma^3}{D} t^0 - \frac{\Delta + \Sigma^1}{D} t^2 \right] \tag{21}$$

$$t^1 = \int \frac{d^2 p}{(2\pi)^2} v \left[ \frac{\omega - \Sigma^0}{D} t^2 + \frac{\varepsilon + \Sigma^3}{D} t^1 - \frac{\Delta + \Sigma^1}{D} t^3 \right] \tag{22}$$

$$t^2 = \int \frac{d^2 p}{(2\pi)^2} v \left[ \frac{\omega - \Sigma^0}{D} t^1 + \frac{\varepsilon + \Sigma^3}{D} t^2 + \frac{\Delta + \Sigma^1}{D} t^0 \right] \tag{23}$$

$$t^3 = v + \int \frac{d^2 p}{(2\pi)^2} v \left[ \frac{\omega - \Sigma^0}{D} t^0 + \frac{\varepsilon + \Sigma^3}{D} t^3 + \frac{\Delta + \Sigma^1}{D} t^1 \right] \tag{24}$$

with

$$D(\mathbf{p}, \omega) = \left( \omega - \Sigma^0 \right)^2 - \left( \varepsilon + \Sigma^3 \right)^2 - \left( \Delta + \Sigma^1 \right)^2 \tag{25}$$

These equations are obviously very difficult to solve in all generality and nobody has yet succeeded in deriving a complete solution! They have been solved by a large number of authors, including the present ones, (Hensen et al., 1997; Rieck et al., 2005; Scheffler, 2004; Balatsky et al., 2006) using a variety of approximations. We shall show here that some approximations, while leading to seemingly reasonable results, cannot be trusted.

## 3. Selfconsistent $\hat{T}$–matrix approximation (SCTMA) in the Fermi surface restricted approximation

In the quasiclassical approximation, the energy integration is performed assuming particle-hole symmetry. Terms in Eqs. (21) - (24) with $\varepsilon + \Sigma^3$ in the numerator, which are responsible for the divergence of the real part of $\hat{G}^0(0;\omega)$, then vanish. This quasiclassical approximation is justified (Eilenberger, 1968; Larkin and Ovchinnikov, 1969) with the argument that only differences between the superconducting and the normal state need to be considered. In Eqs. (21) - (24), subtracting the corresponding normal state equations does not seem to improve the convergence of the $\varepsilon$-integral.

In the present case of a translationally invariant system, the quasiclassical approximation reduces to the omission of terms odd in energy, putting momenta equal to their values on the Fermi surface (line) except when they appear as argument of $\varepsilon$, and then integrating with respect to the $\varepsilon$-dependence of the Denominator (25). For a circular Fermi surface all momenta are of the form $k_F(\cos\varphi, \sin\varphi)$ so that the set of equations (21) - (24) reduces to

$$t^0(\varphi,\phi) = \pi N_F \int_0^{2\pi} \frac{d\psi}{2\pi} v(\varphi-\psi) \left[g^0(\psi) t^3(\psi,\phi) - g^1(\psi) t^2(\psi,\phi)\right] \quad (26)$$

$$t^1(\varphi,\phi) = \pi N_F \int_0^{2\pi} \frac{d\psi}{2\pi} v(\varphi-\psi) \left[g^0(\psi) t^2(\psi,\phi) - g^1(\psi) t^3(\psi,\phi)\right] \quad (27)$$

$$t^2(\varphi,\phi) = \pi N_F \int_0^{2\pi} \frac{d\psi}{2\pi} v(\varphi-\psi) \left[g^0(\psi) t^1(\psi,\phi) + g^1(\psi) t^0(\psi,\phi)\right] \quad (28)$$

$$t^3(\varphi,\phi) = v(\varphi-\phi) + \pi N_F \int_0^{2\pi} \frac{d\psi}{2\pi} v(\varphi-\psi) \left[g^0 t^0(\psi,\phi) + g^1 t^1(\psi,\phi)\right] \quad (29)$$

$g^0(\psi,\omega_+)$ and $g^1(\psi,\omega_+)$ are the energy integrated normal and anomalous retarded Green functions

$$g^0(\psi,\omega_+) = -\frac{\omega - \Sigma^0(\psi,\omega_+)}{\sqrt{[\Delta(\psi) + \Sigma^1(\psi,\omega_+)]^2 - [\omega - \Sigma^0(\psi,\omega_+)]^2}} \quad (30)$$

$$g^1(\psi,\omega_+) = -\frac{\Delta(\psi) + \Sigma^1(\psi,\omega_+)}{\sqrt{[\Delta(\psi) + \Sigma^1(\psi,\omega_+)]^2 - [\omega - \Sigma^0(\psi,\omega_+)]^2}} \quad (31)$$

Since particle-hole symmetry is assumed, $g^0$ and $g^1$ are independent of $t^3(\psi,\psi)$ and $g^3$ vanishes. However, all four components $t^\ell(\varphi,\phi)$ are required for the calculation of $\Sigma_{0,1}$.

If the defect potentials are taken to be $\delta$-functions in real space, $v = v_0$ is independent of angle and so are the $t^\ell$. Then one has to take the Fermi surface average of $g^1(\psi)$, which vanishes. Hence $t^1 = t^2 = 0$ and

$$t^0 = \frac{\pi N_F v_0^2 <g^0>}{1 - (\pi N_F v_0)^2 <g^0>^2}, \quad t^3 = \frac{v_0}{1 - (\pi N_F v_0)^2 <g^0>^2} \quad (32)$$

for arbitrarily large $v_0$. These are standard results derived and used by many authors, including the present ones (Hensen et al., 1997), to describe (transport) properties of cuprate superconductors.

Here, we shall show that a straightforward extension of this theory to defects with finite range leads to unacceptable results!

For specific examples of impurity potentials with finite range we consider a Gaussian and a disk:

$$v_G(r) = \bar{v}\frac{1}{\pi a^2} e^{-r^2/a^2} \quad v_D(r) = \bar{v}\begin{cases} \frac{1}{\pi a^2} & \text{for} \quad r < a \\ 0 & \text{for} \quad r > a \end{cases} \quad (33)$$

Both potentials reduce to $\delta$-functions when the range $a$ goes to zero while the spatial average $\bar{v} = v_{\max}\pi a^2$ is kept constant. Fourier transformation gives:

$$v_G(\mathbf{k}' - \mathbf{k}) = \bar{v}\, e^{-(\mathbf{k}'-\mathbf{k})^2 a^2/4} \quad v_D(\mathbf{k}' - \mathbf{k}) = \bar{v}\frac{2}{|\mathbf{k}' - \mathbf{k}|a} J_1(|\mathbf{k}' - \mathbf{k}|a) \quad (34)$$

Because of the symmetry assumed for the normal state electronic structure, both functions can be expanded into a cosine series with respect to the scattering angle $\varphi$ between $\mathbf{k}$ and $\mathbf{k}'$. For brevity, we write the series in complex form:

$$v(k', k, \cos\varphi) = \sum_{n=-\infty}^{+\infty} v_n(k', k)\, e^{in\varphi} \quad \text{with} \quad v_n(k', k) = \bar{v}e^{-\frac{1}{4}(k'^2+k^2)a^2} I_n\left(\frac{k'ka^2}{2}\right), \quad (35)$$

where the explicit expression for $v_n(k', k)$ applies to the Gaussian potential. The $I_n$'s are modified Bessel functions. For the disk-shaped potential, these coefficients have to be evaluated numerically.

Within the Fermi surface restricted approach, $k' = k = k_F$. Hence, it is convenient to redefine the above expansion in terms of new parameters:

$$v(k_F, \cos\varphi) = v_0 \sum_{n=-\infty}^{+\infty} u_n e^{in\varphi} \quad \text{with} \quad (36)$$

$$v_0 = \bar{v}e^{-\frac{1}{2}k_F^2 a^2} I_0(\gamma), \quad u_n = \frac{I_n(\gamma)}{I_0(\gamma)} \quad \text{and} \quad \gamma = \frac{1}{2}k_F^2 a^2. \quad (37)$$

These expansion coefficients are often treated as free parameters.

Just like the defect potential the $t^\ell$'s are expanded into Fourier series' and the coefficients are collected in the form of matrices $\tilde{t}^\ell$ with elements:

$$t^\ell_{nm} = \frac{\pi N_F}{\sin^2 \delta_s} \int_0^{2\pi} \frac{d\varphi}{2\pi} \int_0^{2\pi} \frac{d\phi}{2\pi} t^\ell(\varphi, \phi) e^{-in\varphi + im\phi}. \tag{38}$$

For later convenience we have defined $t^\ell_{nm}$ with a factor depending on an *s*-wave scattering phase shift $\delta_s$, which characterizes the strength of the potential (37)

$$\pi N_F v_0 = \tan \delta_s. \tag{39}$$

It follows from symmetry consideration (Klemm et al., 2000) that the most general real order parameter transforming as $k_x^2 - k_y^2$ can be written in a Fermi surface restricted approach as

$$\Delta(T, \varphi) = \sum_{m=-\infty}^{\infty} a_{4m-2} \cos[(4m-2)\varphi]. \tag{40}$$

Then $g^\ell(\psi; \omega)$ has the general form

$$g^\ell(\psi; \omega) = \sum_{m=-\infty}^{+\infty} g^\ell_{q,q+|4m-2\ell|}(\omega) \cos[(4m-2\ell)\psi] \qquad \ell = 0, 1. \tag{41}$$

It is convenient to write the expansion coefficients in the form of symmetric matrices $\tilde{g}^\ell = (g^\ell_{q,k})$ with the additional property $g^\ell_{q,k} = g^\ell_{-q,-k}$, even though $g^\ell_{q,q+m}$ is actually independent of $q$. Since the matrix elements are complex, these matrices are not hermitian. When the Fourier coefficients $u_n$ of the potential (35) are written in the form of a diagonal matrix $\tilde{u}$, the four integral equations (26) - (29) are transformed to four equations for $\tilde{t}^\ell$. From these, $\tilde{t}^2$ and $\tilde{t}^3$ can be eliminated so that we finally obtain

$$\tilde{t}^0 \pm \tilde{t}^1 = \left[\cos^2\delta_s - \sin^2\delta_s \, \tilde{u} \, (\tilde{g}^0 \mp \tilde{g}^1) \, \tilde{u} \, (\tilde{g}^0 \pm \tilde{g}^1)\right]^{-1} \tilde{u} \, (\tilde{g}^0 \mp \tilde{g}^1) \, \tilde{u}. \tag{42}$$

This is the central equation that has to be solved numerically for various choices of the scattering potential (35). Because of the selfconsistency requirement the solution depends also on $n_{\text{imp}}$ through the parameter $\Gamma_N^{\text{el}}$ (49).

The $t^\ell(\varphi, \phi)$ have the following symmetries

$$t^{0,1,3}(\varphi, \phi) = t^{0,1,3}(\phi, \varphi); \quad t^2(\varphi, \phi) = -t^2(\phi, \varphi); \quad t^\ell(\varphi, \phi) = t^\ell(-\varphi, -\phi), \tag{43}$$

from which we derive

$$t^{0,1}_{m,n} = t^{0,1}_{n,m} = t^{0,1}_{-m,-n}. \tag{44}$$

It follows from (41) and (42) that

$$t^0_{q,q'} = t^0_{q,q+4m}\, \delta_{q',q+4m} \tag{45}$$

$$t^1_{q,q'} = t^1_{q,q+4m-2}\, \delta_{q',q+4m-2} \tag{46}$$

In contrast to (41), $t^{0,1}_{q,q'}$ does depend on $q$. Hence, the selfenergy (19) has the same structure (41) as $g^{0,1}(\psi;\omega)$:

$$\Sigma^1(\psi;\omega) = \Gamma^{\text{el}}_N \sum_{\ell=-\infty}^{+\infty} \left\{ \sum_q t^1_{q,q+4\ell-2} \right\} \cos[(4\ell-2)\psi] \equiv \sum_{\ell=-\infty}^{+\infty} \Sigma^{1\ell} \cos[(4\ell-2)\psi] \tag{47}$$

$$\Sigma^0(\psi;\omega) = \Gamma^{\text{el}}_N \sum_{\ell=-\infty}^{+\infty} \left\{ \sum_q t^0_{q,q+4\ell} \right\} \cos[4\ell\psi] \equiv \sum_{\ell=-\infty}^{+\infty} \Sigma^{0\ell} \cos[4\ell\psi] \tag{48}$$

where the parameter

$$\Gamma^{\text{el}}_N = \frac{n_{\text{imp}}}{\pi N_F} \sin^2 \delta_s \tag{49}$$

has been introduced.

3.1. RESULTS FOR SELFENERGIES AND PAIR BREAKING PARAMETERS

Results for the selfenergies $\Sigma^{00}$ and $\Sigma^{01}$ are shown in Figure (1) for a Gaussian potential with $\gamma = 5$. The OP has been taken to be $\Delta(T;\varphi) = \Delta_{\max}(T)\cos 2\varphi$ with a low temperature amplitude $\Delta_{\max} = 16\,\text{meV}$. The exact numerical value is of no physical significance here. We focus our attention on the limiting behavior for $\omega \gg \Delta_{\max}$. In this limit, $\Sigma^{0\ell}$ reduces to the normal state result: $\Sigma^{0\ell} = 0$ for $\ell \neq 0$ because in the system considered, rotational symmetry is broken only by the superconducting order parameter. The purely imaginary isotropic contribution

$$\Sigma^{00}(\omega_+) = -i\Gamma^{\text{el}}_N \sum_{\ell=-\infty}^{\infty} \frac{u^2_\ell}{\cos^2 \delta_s + \sin^2 \delta_s\, u^2_\ell} \tag{50}$$

is easily obtained because the Green function (30) reduces to $g^0(\psi,\omega_+) = -i$ while $\tilde{t}^0$ in (42) is diagonal and both $\tilde{t}^1$ and $\tilde{g}^1$ can be neglected. For $\delta$-function potentials $u_\ell = 0$ for $\ell \neq 0$ and (50) reduces to $\Sigma^{00}(\omega_+) = -i\Gamma^{\text{el}}_N$, which elucidates the physical significance of the parameter $\Gamma^{\text{el}}_N$ introduced above. In the Born limit $\delta_s \ll \pi/2$, the denominator in (50) can be replaced by 1. The remaining sum is related to the Fermi surface average of the squared potential: $\langle v^2(\varphi)\rangle = \langle v(\varphi)\rangle^2 \sum_{\ell=-\infty}^{\infty} u^2_\ell$. For a Gaussian potential (37) the sum can easily be performed yielding $\Sigma^{00}(\omega_+) = -i\Gamma^{\text{el}}_N I_0(2\gamma)/I_0^2(\gamma)$. For $\gamma = 5$, the correction factor is 3.8. In the unitary limit $\delta_s = 0.5\pi$ the sum in (50) diverges because $u^2_\ell$ cancels, no matter how small. The numerical results shown in Fig. (1) were obtained with a cut-off $\ell_{max} = 10$, hence the limiting value of 4.2 meV. Whether the set of $u_\ell$'s represent a Gaussian or any other potential is obviously immaterial for this argument.

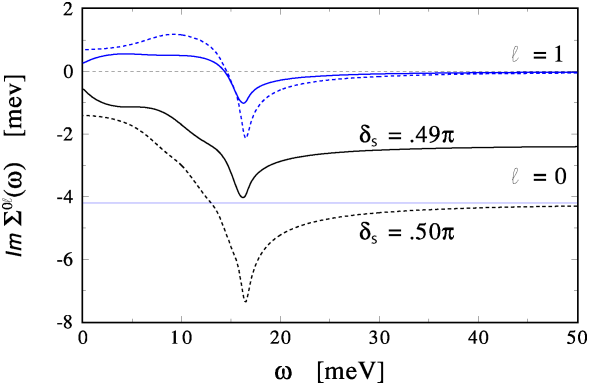

*Figure 1.* Imaginary part of the selfenergy defined in (48) for Gaussian defect potentials (34) with $\gamma = 5$ (37) as function of frequency for $\Gamma_N^{el} = 0.2$ meV and defect potentials at or near the unitary limit. The parameter $\delta_s$ is defined in (39) Extrema occur at $\omega = \Delta_{max}$.

One might argue that an infinitely high potential is unphysical. However, as known from elementary quantum mechanics, a hard disk or sphere simply presents a part of space which the particles cannot enter and the difference between a large and an infinite potential is, in fact, not substantial. In Fig. (1) we also show results for $\delta_s = 0.49\pi$. In this case, contributions from terms $\ell > 7$ are negligible. However, the limiting value is still much larger than for point-like scatterers or weak scatterers. This appears to be an artefact of the Fermi surface restricted approximation, which goes unnoticed when one starts with considering scattering only in very few angular momentum channels.

This theory has been used by us to calculate the $T_c$-degradation of an unconventional superconductor with an order parameter of the general form (40) (Rieck et al., 2005). In this context one requires $\Sigma^1(\psi, i\omega_n)$ which also diverges for $\delta_s = \pi/2$. For one component order parameters $\Delta(\varphi) = \Delta_{max} \cos(4\ell - 2)\varphi$ we find the standard Abrikosov-Gor'kov formula

$$\ln \frac{T_c}{T_{c0}} = \psi\left(\frac{1}{2}\right) - \psi\left(\frac{1}{2} + \frac{\lambda_{4\ell-2}}{2}\right) \quad (51)$$

with pair breaking parameters

$$\lambda_{4\ell-2} = \frac{\Gamma_N^{el}}{\pi T_c} \frac{1}{2} \sum_{m=-\infty}^{\infty} \frac{(u_m - u_{m+4\ell-2})^2}{(\cos^2 \delta_s + u_m^2 \sin^2 \delta_s)(1 + u_{m+4\ell-2}^2 \tan^2 \delta_s)} \quad (52)$$

Taking the unitary limit in (52) it would appear as if every term in the series vanishes. However, when the series is terminated at $m = \pm m_0$ (Kulić and Dolgov, 1999), one finds $\lambda_{4\ell-2} = \frac{1}{\pi T_c} \Gamma_N^{el}(4\ell - 2)$, i.e. 2 and 6 for the examples shown in Fig. (2). independent of $m_0$. These results are clearly also artefacts of the Fermi surface restricted approximation and have nothing to do with physical reality.

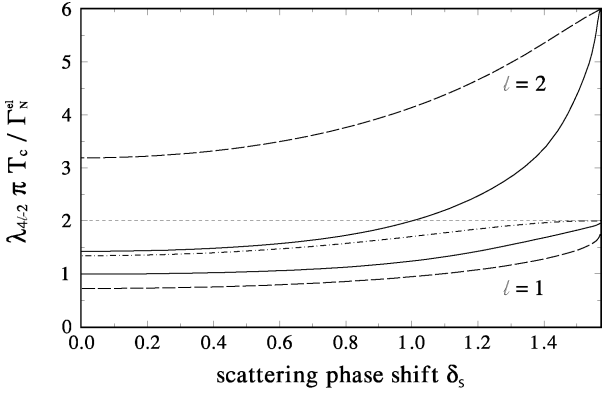

Figure 2. $\lambda_2$ and $\lambda_6$ for Gaussian potentials with widths $\gamma = 1$ (solid) and $\gamma = 5$ (dashed) as function of potential strength, parametrized by $\delta_s = \tan^{-1}(\pi N_F v_0)$. The dot-dashed line shows the result for $\lambda_2$ when only $s$-, $p$-, and $d$-wave scattering is included. For $\delta$-function scatterers the reduced pairbreaking parameters plotted in this figure are identically 1.

## 4. Single spherically symmetric impurity in the normal state

In order to elucidate the reason for the failure of the Fermi surface restricted approach, we investigate scattering off a single spherically symmetric impurity in the normal state. To make contact with the scattering phase shift analysis known from elementary quantum mechanics we introduce the scattering wave functions $\psi_k(r)$. From these one can construct the Green function

$$G(r, r', \omega_+) = \int \frac{d^2k}{(2\pi)^2} \frac{\psi_k(r) \psi_k^*(r')}{\omega_+ - \varepsilon(k)} \tag{53}$$

which fulfills the integral equation (5), suitably simplified to the normal state. Inserting this representation for $G$ into (5) and projecting out the $r'$-dependence, one can neglect the inhomogeneous term when the limit $\omega \to \varepsilon(k)$ is taken. One thus obtains an equation for $\psi_k(r)$:

$$\psi_k(r) = e^{ik \cdot r} + \int d^2\rho\, G^0(r - \rho; \varepsilon(k))\, V(\rho) \psi_k(\rho). \tag{54}$$

This equation could, of course, also have been derived directly from the time-independent Schrödinger equation.

Inserting the plane-wave representation of $G^0(r - \rho; \varepsilon(k))$ and defining a $T$-matrix through

$$T(p, k) = \int d^2r\, e^{ip \cdot r}\, V(r) \psi_k(r), \tag{55}$$

we arrive at

$$\psi_k(r) = e^{ik \cdot r} + \int \frac{d^2p}{(2\pi)^2} e^{ip \cdot r}\, \hat{G}^0(p, \varepsilon(k))\, T(p, k) \tag{56}$$

The Lippman-Schwinger equation is obtained by multiplying (56) by $e^{-i\mathbf{k}\cdot\mathbf{r}} V(\mathbf{r})$ and then integrating with respect to $d^2r$:

$$T(\mathbf{k},\mathbf{k}') = v(\mathbf{k}-\mathbf{k}') + \int \frac{d^2p}{(2\pi)^2} v(\mathbf{k}-\mathbf{p}) G^0(\mathbf{p},\varepsilon(\mathbf{k}')) T(\mathbf{p},\mathbf{k}') \quad (57)$$

This is identical with (15) when the energy variable $\omega$ in the generalized $T$-matrix is replaced by $\varepsilon(\mathbf{k})$ (Mahan, 1981). Solving (57) can be simplified if only a single defect, described by a spherically symmetric potential $v(r)$ is considered. Then the $T$-matrix depends on the moduli of $\mathbf{p}$ and $\mathbf{k}$ and the angle between them, even when these vectors belong to different energies. Eq.(57) can thus be rewritten as

$$T(k',k,\cos\phi) = v(k',k,\cos\phi) +$$
$$+ \int_0^\infty \frac{dp\, p}{2\pi} \int_0^{2\pi} \frac{d\theta}{2\pi} v(k',p,\cos(\phi-\theta)) G^0(p,\varepsilon(k)) T(p,k,\cos\theta) \quad (58)$$

For an isotropic system the expansion of $T$ into a Fourier series analogous to (35), leads to a set of decoupled 1D integral equations

$$T_m(k',k) = v_m(k',k) + \int_0^\infty \frac{dp\, p}{2\pi} v_m(k',p) G^0(p,\varepsilon(k)) T_m(p,k). \quad (59)$$

When the local density of states near an impurity is the quantity of interest, it is a reasonable approximation to use the Green function for the clean system $G^0(p,\varepsilon(k)) = \left[\frac{k^2}{2\mu} - \frac{p^2}{2\mu} + i\delta\right]^{-1}$ so that we can rewrite (59) as

$$T_m(k',k) = v_m(k',k) +$$
$$+ \frac{\mu}{\pi}\mathcal{P}\int_0^\infty dp \frac{p}{k^2-p^2} v_m(k',p)T_m(p,k) - i\frac{\mu}{2} v_m(k',k)T_m(k,k) \quad (60)$$

Assuming cylindrically symmetric impurities, the $T$-matrix equation in the alloy model (17) can be reduced to an equation identical to (59), except that the Green function is to be replaced by

$$G(p,\omega_+) = \left[\omega - \frac{p^2}{2\mu} - \Sigma(p,\omega_+)\right]^{-1}. \quad (61)$$

The pole of $G^0(p,\omega_+)$ has moved into the complex plane so that the separation into a principal value part and a $\delta$-function is not possible. Furthermore, the $T$-matrix has to be calculated as function of the parameter $\omega$ in order to obtain the selfenergy $\Sigma(p,\omega_+)$ (19), which could then be compared with the high frequency limit in Fig. 1. This work is in progress.

We are interested in strong (repulsive) scattering potentials because they can create states inside energy gaps. The strength of the scattering potential can be arbitrary, even a hard sphere should pose no problem! It seems that one could simplify equation (59) and (60) in the case of very strong potentials, because then the $T$-matrix on the left hand sides could be neglected. One thus arrives at a Fredholm integral equation of the first kind which, however, represents an ill-posed problem. The only numerical method for solving this equation that we have found feasible was a discretization, keeping the potential finite.

In order to calculate the wave function from (60), we expand the $T$-matrix (58) and both exponentials in (56) using

$$e^{ikr\cos\varphi} = J_0(kr) + 2\sum_{m=1}^{\infty} i^m J_m(kr) \cos m\varphi \tag{62}$$

and obtain

$$\psi_k(r) = \sum_{m=-\infty}^{\infty} i^m \left\{ J_m(kr) + \int_0^{\infty} \frac{dp\, p}{2\pi} J_m(pr) G_k^0(p) T_m(p,k) \right\} e^{im\varphi}. \tag{63}$$

### 4.1. SCATTERING PHASE SHIFT ANALYSIS

We want to establish the relations between the Fourier coefficients of the $T$-matrix and the partial scattering phase shifts. These relations put some constraints on the $T$-matrix on the energy shell which serve as a useful check for the numerical calculations. The desired relations are derived by considering the asymptotic behavior of the scattered wave function (54).

The normal state Green function is obtained from $G^{00}(r, \omega_+) + G^{03}(r, \omega_+)$ in the limit $\Delta \to 0$ (Eqs. (11,13)):

$$G^0(r, \omega_+) = -\frac{\mu}{2}(iJ_0(kr) - Y_0(kr)) \quad \text{with} \quad k^2 = k_F^2 + 2\mu\omega. \tag{64}$$

According to (73) we immediately obtain the density of states per spin (per area)

$$\mathcal{N}(E(k)) = \frac{\mu}{2\pi}. \tag{65}$$

Since this is independent of energy, it is identical with the parameter $N_F$ introduced in Section 3. Using the asymptotic expansions of the Bessel functions we find:

$$G^0(r - \rho, \omega_+) \asymp -i\, e^{-\frac{i}{4}\pi} \frac{\mu}{\sqrt{2\pi}} \frac{1}{\sqrt{k|r-\rho|}} e^{ik|r-\rho|}. \tag{66}$$

When this is inserted into (54) we can use the approximation $k|r - \rho| \approx kr$ in the denominator and $k|r - \rho| \approx kr - k' \cdot \rho$ with $k' = ke_r$ in the exponent, assuming

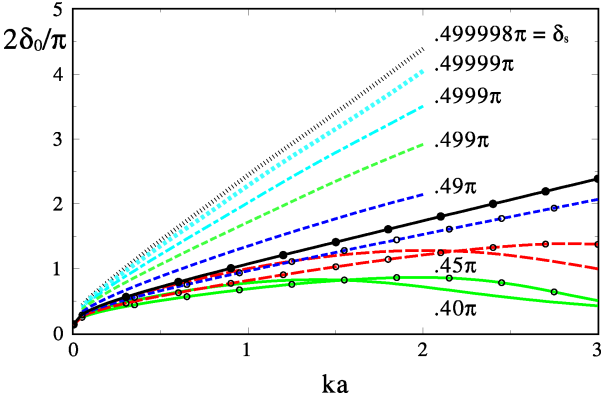

*Figure 3.* s-wave scattering phase shifts as functions of $ka$. Curves marked with circles are the results for a disk, the unmarked ones are for a Gaussian potential. The parameters are $\tan \delta_s = \pi \mathcal{N} \bar{v}$, characterizing the strengths of the scattering potential by its spatial average (33). For $a \to 0$, $\delta_0$ and hence the $T$-matrix vanish logarithmically!

that the potential $v(r)$ is short ranged. Thus, the behavior of the scattered wave function at large distances from the scattering center is given by:

$$\psi_k(r) \asymp e^{i k \cdot r} - i e^{-\frac{i}{4}\pi} \frac{\mu}{\sqrt{2\pi}} \frac{e^{ikr}}{\sqrt{kr}} T(k', k). \tag{67}$$

This involves the $T$-matrix on the energy shell $|k'| = |k| = \sqrt{2\mu E}$ only. Because we started from the retarded Green function, there is only an outgoing cylindrical wave. The prefactor $1/\sqrt{r}$ is required for normalization in two dimensions. Note that this wave function decays even more slowly than in three dimensions which renders the assumption usually made with regard to the independence of scattering events at different defects rather questionable.

Using (62), with the asymptotic expansions of the Bessel functions inserted, we expand (67) in a cosine series, which we again write in complex form to save space:

$$\psi_k(r) \asymp \frac{1}{\sqrt{2\pi kr}} e^{-\frac{i}{4}\pi} \sum_{m=-\infty}^{\infty} \left( (-1)^m i e^{-ikr} + (1 - i\mu T_m(k, k)) e^{+ikr} \right) e^{im\phi} \tag{68}$$

Particle conservation implies that the currents represented by the incoming and the outgoing waves should cancel. Hence, the factor multiplying $e^{+ikr}$ in (68) can only be a phase factor which is usually written in the form

$$1 - i\mu T_m(k, k) = e^{+2i\delta_m} \quad \text{or} \quad \frac{\mu}{2} T_m(k, k) = \frac{1}{\cot \delta_m + i}. \tag{69}$$

This defines the scattering phase shift $\delta_m$ for the $m$'th angular momentum channel. The complex quantity $T_m(k, k)$, which is obtained by numerically solving Eq. (59), can thus be expressed in terms of one real quantity. This provides us with an excellent check for our numerical calculations! Note that no such relation exists for the generalized *T*-matrix.

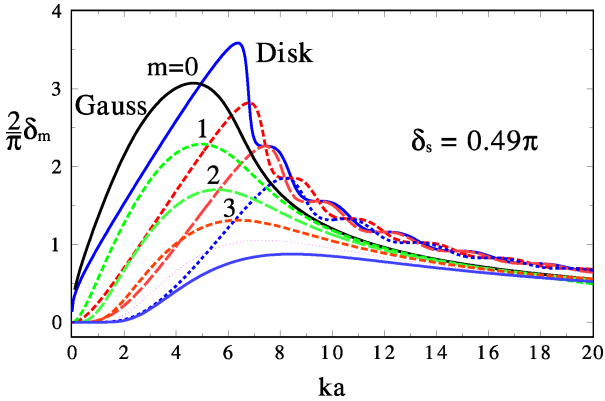

Figure 4. *s*-, *p*-, *d*-, and *f*-wave scattering phase shifts for a Gaussian and a disk-shaped potential (33) as function of *ka*. The range of *ka* is much larger than in Fig. 3.

Results for the *s*-wave scattering phase shift for varying potential strengths $\tan\delta_s = \pi\mathcal{N}\bar{v}$ are shown in Fig. 3. For disk-shape potentials the change in $\delta_0$ when the potential strength $\bar{v}$ is increased from some large value to infinity is small. For the Gaussian potential the results are very similar as long as the potential is fairly weak. However, $\delta_0$ increases indefinitely with $\bar{v}$. This is due to the fact that any potential with smoothly decreasing tails will be infinite everywhere in space when the maximum goes to infinity. We note that $\delta_0 \to 0$ when the potential range *a* goes to zero, no matter how strong the potential. To get resonant scattering ($2\delta_0/\pi = 1$) one does require a rather large potential, but it will occur only for $ka \approx 1$, i.e. when the wavelength of the scattered particle is comparable with the range of the potential.

In Fig. 4 scattering phase shifts for various angular momentum channels are shown for a fairly strong potential and a wide range of *ka*. As expected, $\delta_m$ goes to zero for $a \to 0$ more and more rapidly as *m* increases. Nonetheless, resonance occurs in all *m*-channels considered, albeit at very different values of *ka*. For $ka \gg 1$, the scattering phase shifts become almost independent of *m*!

Results for the disk-shaped potential shown in these two figures have been calculated from (60) which is very time consuming because integrals have to be evaluated to obtain the $v_m(k',k)$'s. Our numerical results agree perfectly with

$$\tan\delta_m(k) = -\frac{qaI'_m(qa)J_m(ka) - kaI_m(qa)J'_m(ka)}{qaI'_m(qa)Y_m(ka) - kaI_m(qa)Y'_m(ka)} \quad \text{for} \quad v_{\max} \geq \frac{k^2}{2\mu}, \quad (70)$$

$$\text{with} \quad qa = \sqrt{\frac{4}{\pi}\tan\delta_s - (ka)^2} \geq 0, \quad (71)$$

which is derived by solving the Schrödinger equation for $r < a$ and for $r > a$ and then matching the logarithmic derivatives of the wave functions at $r = a$.

This comparison shows that it is absolutely essential to keep the principle value integral in (60). When this is neglected, as is the case in the Fermi surface restricted approach, one immediately finds

$$\tan \delta_m(k) = \pi \mathcal{N} v_m(k,k) \leq \tan \delta_s. \tag{72}$$

Hence, resonant scattering occurs only for $\bar{v} \to \infty$. Now, however, all angular momentum channels become resonant simultaneously. This is not only unphysical, it also causes convergence problems with the Fourier expansion of (58)!

## 4.2. RESULTS FOR THE LOCAL DENSITY OF STATES

From (60) and (63) we calulated the Green function (53) and, in particular, the Local Density of States (LDOS), which is experimentally accessible through Scanning Tunneling Microscopy (Crommie et al., 1993):

$$\mathcal{N}(r,E) = -\frac{1}{\pi} \mathcal{I}m\, G_E(r,r) = \int \frac{d^2k}{(2\pi)^2} |\psi_k(r)|^2 \delta\left(E - \frac{k^2}{2\mu}\right). \tag{73}$$

The result is rotationally invariant, as expected. Note that, in order to calculate $\mathcal{N}(r,E)$ at arbitrary distances from the scattering center, we need $T_m(p,k)$ off the energy shell. In the asymtotic regime we can use (68) to calculate the LDOS. For the disk-shaped potential this is almost trivial because we have the analytic results (70) for $T_m(k,k)$. In Fig. (5) we included results for the LDOS, normalized to the bulk DOS $\mathcal{N}$, which are valid only in the asymptotic regime and compare them with one exact result. The most remarkable conclusion, in view of the discussions in the literature centered around $\delta$-function scatterers, to be drawn from this figure is that nothing exceptional happens when the scattering becomes resonant. This can be attributed to an overdamping of resonant states due to the bulk density of states being finite at all energies. This can be seen more formally from Eq (69) because this gives

$$\pi \mathcal{N} |T_m(k,k)| \leq 1. \tag{74}$$

Well-defined resonant states can only be expected inside energy gaps of the bulk DOS. It seems to be rather difficult, though, to devise a consistent toy model to study such effects. (Joynt, 1997) The most appropriate applications that come to mind are unconventional superconductors, which we are working on.

Since we have varied the energy of the scattered particle rather than the potential, the wavelength of these Friedel oscillations changes. Also remarkable is the amplitude of these oscillations which is much greater than in three dimensions (Fetter, 1965). The slow decay of the oscillation is related to the two-dimensionality of the system considered.

In Fig. 6 corresponding exact results are shown for a Gaussian potential. For the same potential strength $\pi \mathcal{N} \bar{v} = 31.82$, the results for the two model potentials are very similar. For a very high potential, $\delta_s = 0.49999\pi$, the LDOS vanishes at distances much larger than the potential range. This is related to the peculiar behavior of the scattering phase shift discussed in connection with Fig. 3.

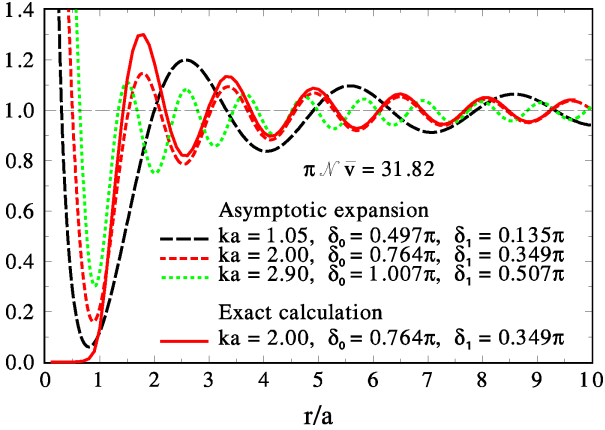

*Figure 5.* Local density of states as function of distance for a disk-shaped potential (33) with radius $a$ centered at the origin. A very high value $\pi \mathcal{N} \bar{v} = \tan \delta_s = 31.82 \Leftrightarrow \delta_s = 0.49\pi$ has been assumed for the potential. Some curves have been calculated using the asymptotic expansion (68) rather than (63) The values of $ka$ have been chosen such that either $s$-wave ($\delta_0 = 0.5\pi$) or $p$-wave ($\delta_1 = 0.5\pi$) scattering or neither of them are resonant. For $ka = 2.00$, exact results can be compared with results obtained using the asymptotic form of the wave function.

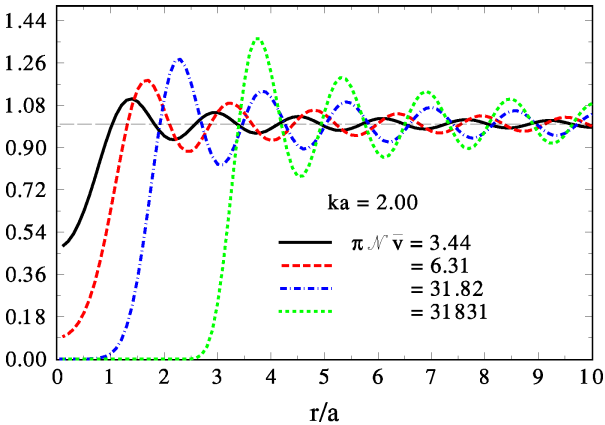

*Figure 6.* Local density of states as function of distance for a Gaussian potential (33) with range $a$ centered at the origin.

The Green function, and with it $\mathcal{N}(r, E)$, could also be obtained from Eqs. (14) and (15), where $E = \omega + \epsilon_F$. Then one has to replace $\varepsilon(k)$ in (59) by $\omega$ and solve this set of equations for a range of values of this new variable. Introducing Fourier expansions, integrals with respect to angles in (14) can be done, rendering $G(r, r, \omega)$ isotropic. We are then left with a double integral, while the calculation of the LDOS from (73) only involves a set of one-dimensional integrals. Assuming $\delta$-function scatterers, these equations immediately lead back to Eq. (9).

## 5. Summary

We have shown that an approximate treatment of potential scattering in a metal (superconductor), which puts momentum variables on the Fermi surface and evaluates the energy-integrated Green function invoking particle-hole symmetry, leads to unacceptable results when the potentials have finite ranges. The key problem is that in the strong scattering limit results will depend on the number of angular momentum channels taken into account. When only a single scattering channel (*s*-wave scattering, $\delta$-function potential) or very few scattering channels are considered, this problem is not apparent. Except, of course, for the discrepancy that elementary quantum mechanics tells us that a $\delta$-function potential does not scatter in two dimensions.

The only way to obtain acceptable results and to reproduce standard results within a continuum description is to take all momentum and energy dependencies accurately into account. This involves the evaluation of principle value integrals and, in the most general cases, the numerical solution of two-dimensional Fredholm integral equations of the second kind. In the strong scattering limit these go over into those of the first kind. However, these cannot be solved numerically without some regularization. One such regularization consists of reverting back to an equation of the second kind. For a disk-shape potential we have shown that the results for a strong potential and an infinite potential do not significantly differ, as one would expect. For potentials that go to zero continuously, like Gaussians or Lorentzians, it has turned out to be unphysical to let the height of the potential to go to infinity.

Comparison with alternative approaches based on a localized description of the material is an important project for the future.

## Acknowledgements

The authors gratefully acknowledge useful discussions with A. V. Balatsky, J. C. Davis, A. Lichtenstein, N. Schopohl, S. Trugman, and J.-X. Zhu. One of the authors (K.S.) is very grateful for the hospitality extended to him at the Los Alamos National Laboratory, where part of this work was done.